\documentclass[aps,prb,showpacs,floatfix]{revtex4}
\usepackage[dvips]{color}
\usepackage{graphicx}
\usepackage{amssymb}
\usepackage{rotating}
\usepackage{epsfig}

\newlength{\TZ}
\newcommand{\BEQ}{\begin{equation}}     
\newcommand{\BEA}{\begin{eqnarray}}
\newcommand{\EEQ}{\end{equation}}       
\newcommand{\EEA}{\end{eqnarray}}
\newcommand{\II}{{\rm i}}               

\newcommand{\appsection}[2]{\setcounter{equation}{0}\setcounter{subsection}{0}
\section*{Appendix #1. #2}
\renewcommand{\theequation}{#1\arabic{equation}}
              \renewcommand{\thesection}{#1} }

\catcode`\@=11
\def\numberbysection{\@addtoreset{equation}{section}
        \def\theequation{\thesection.\arabic{equation}}}

\begin{document}
\input epsf.sty

\title[Local aging phenomena close to magnetic surfaces]{Local
aging phenomena close to magnetic surfaces}

\author{Florian Baumann$^{1,2}$ and 
Michel Pleimling$^3$} 
\affiliation{$^1$Laboratoire de Physique des 
Mat\'eriaux\footnote{Laboratoire associ\'e au CNRS UMR 7556},
Universit\'e Henri Poincar\'e Nancy I, \\ 
B.P. 239, F -- 54506 Vand{\oe}uvre l\`es Nancy Cedex, France}
\address{$^2$Institut f\"ur Theoretische Physik I, 
Universit\"at Erlangen-N\"urnberg, \\
Staudtstra{\ss}e 7B3, D -- 91058 Erlangen, Germany}
\address{$^3$ Department of Physics, Virginia Polytechnic
Institute and State University, Blacksburg, VA 24061-0435,
USA}

\begin{abstract}
Surface aging phenomena are discussed for semi-infinite systems prepared in a fully
disordered initial state and then quenched to or below the critical point.
Besides solving exactly the semi-infinite Ising model 
in the limit of large dimensions,
we also present results of an extensive numerical study
of the nonequilibrium dynamical behavior of the two-dimensional semi-infinite Ising model 
undergoing coarsening. The studied models reveal a simple aging behavior where some of the
nonequilibrium surface exponents take on values that differ from their bulk counterparts. For the
two-dimensional semi-infinite Ising model we find that 
the exponent $b_1$, that describes the scaling behavior of the surface autocorrelation,
vanishes. These simulations also reveal the existence
of strong finite-time corrections that to some extent mask the leading scaling behavior of the
studied two-time quantities.
\end{abstract}

\maketitle

\setcounter{footnote}{0}

\section{Introduction}
Intriguing phenomena are observed when bringing simple ferromagnets out of equilibrium
through a temperature quench \cite{Bray94,Cugl02,Calabrese04,Hen07,buch}. Consider a ferromagnetic system 
prepared at high temperatures, i.e. in a disordered and uncorrelated initial state,
that is suddenly quenched onto or below the critical point. If the quench is onto the 
critical point, critical dynamics sets in, yielding a dynamical correlation length that
increases with time. For a quench inside the ordered phase, the formation and growth of well-ordered
domains are observed. Interestingly, these two physically very different cases have in common
that the relevant length scale $L$ (which in the first case is the dynamical correlation length,
whereas in the second case it is given by the typical extent of the ordered domains)
increases as a simple power-law of time:
\begin{equation} \label{eq_growth}
L(t) \sim t^{1/z}.
\end{equation}
Assuming non-conserved dynamics (which is the only dynamics studied in this paper) one finds
for the dynamical exponent $z$ the value 2 below the critical temperature $T_c$, whereas at
$T_c$ the dynamical exponent may take on values slightly larger than 2.

The power-law growth (\ref{eq_growth}) is responsible for many nonequilibrium phenomena
which are usually summarized under the header of {\it simple aging}. Thus, it follows directly
from (\ref{eq_growth}) that two-time quantities, like dynamical correlation and response
functions, display dynamical scaling. Introducing the time- and space-dependent order parameter
$\phi(\vec{r};t)$, the correlation function can be written as
\begin{equation} \label{eq_cor_def}
C(t,s;\vec{r}-\vec{r}\;') = \left< \phi(\vec{r};t) \phi(\vec{r}\;';s) \right>,
\end{equation}
whereas the response function, which measures the response of the order parameter 
at site $\vec{r}$ at time $t$ to an external field $h(\vec{r}\;';s)$
acting on site $\vec{r}\;'$ at time $s$, is defined by
\begin{equation} \label{eq_res_def}
R(t,s;\vec{r}-\vec{r}\;') = \left. \frac{\delta \left< \phi(\vec{r};t) \right>}{\delta h(\vec{r}\;';s)}
\right|_{h=0} ~~~~(t>s).
\end{equation}
In writing these equations we assume spatial translation invariance, as encountered in ferromagnetic bulk systems.
The usually studied autocorrelation function $C(t,s)$ and autoresponse function $R(t,s)$ are obtained by setting
$\vec{r}=\vec{r}\;'$ in Equations (\ref{eq_cor_def}) and (\ref{eq_res_def}), respectively.
In the dynamical scaling regime with $t$, $s$, $t-s \gg \tau_{micro}$, where $\tau_{micro}$ is a
microscopic time scale, these two-time quantities can be cast into a simple 
scaling form. For example, for the autocorrelation and the autoresponse functions we have
\begin{equation} \label{eq_dyn_scal}
C(t,s) = s^{-b} f_C(t/s) ~~ \mbox{and} ~~R(t,s) = s^{-1-a} f_R(t/s),
\end{equation}
where $a$ and $b$ are nonequilibrium exponents, whereas $f_C$ and $f_R$ are scaling functions that for 
large arguments display a power-law decay,
\begin{equation} \label{eq_scal_fct}
f_C(y) \sim y^{-\lambda_C/z}~~ \mbox{and} ~~ f_R(y) \sim y^{-\lambda_R/z},
\end{equation}
with the autocorrelation \cite{Fish88,Huse89} and the autoresponse exponents \cite{Pico02} $\lambda_C$ and $\lambda_R$.
At the critical point,
the nonequilibrium exponents $a$ and $b$ can be expressed by known critical exponents, yielding 
$a=b = (d - 2+\eta)/z$ where $d$ is the dimensionality of the system and $\eta$ is the usual static critical
exponent governing the power-law decay of the spatial correlations at equilibrium. In the ordered phase
one has $b=0$ and $a=1/z$ for systems with exponentially decaying static correlations \cite{Bray94,Henk02a,Lorenz07}. 
At the critical point the autocorrelation exponent
is related to the so-called initial slip exponent \cite{Jans89}. 
In addition, autocorrelation
and autoresponse exponents can be shown to be identical in systems with short-ranged initial correlations and
purely relaxational dynamics \cite{Bray94,Pico04}.

This briefly described simple aging scenario has been studied very intensively in bulk systems,
but for systems bounded by surfaces the investigation of local aging processes
close to surfaces is only at its very beginning\cite{Pleimling04,Calabrese04,Baumann06}. 
The emerging interest in surface aging phenomena
can be related to the increasing importance of surface dominated small systems in research and technology.
Indeed, nonequilibrium processes are deeply affected by the presence of surfaces which can result in changes
in the physical behavior even at macroscopic distances from the surface.
In principle, these surface properties can be studied by X-ray scattering at grazing incidence.

Looking at critical systems bounded by surfaces,  
it is well known that already the static critical behavior at a surface is different from
the bulk critical behavior, yielding new sets of static surface critical exponents 
\cite{Bin83,Die86,Die97,Ple04a}. In fact,
one even encounters different surface universality classes for a given bulk universality class,
depending on the value of the surface couplings or on the existence of a surface field. 
Looking at the dynamics, it has 
to be noted that in the
case of purely diffusive dynamics the dynamical exponent $z$ has the same universal value close to the
surface as inside the bulk \cite{Die83,Rit95,Maj96}. One then expects a surface aging behavior similar to the 
bulk behavior, but with local nonequilibrium exponents and
scaling functions of local two-time functions that differ from the corresponding bulk quantities.

In order to discuss the expected aging phenomenology close to a critical surface in more detail, let us consider 
an idealized semi-infinite lattice in $d$ dimensions where we write the position vector $\vec{r}$ as $\vec{r} = (\vec{x},y)$.
Here $\vec{x}$ is a $(d-1)$-dimensional vector parallel to the surface, whereas $y$ labels the layers perpendicular
to the surface (with $y=1$ being the surface layer). 
With this we obtain the following generalizations for the correlation and response functions:
\begin{eqnarray} \label{eq:c_r_semi}
C(t,s;y,y', \vec{x} - \vec{x}\;') & = & \left< \phi(\vec{x},y;t) \phi(\vec{x}\;',y';s) \right>~, \nonumber \\
R(t,s;y,y', \vec{x} - \vec{x}\;') & = & \left. \frac{\delta \left<
\phi(\vec{x},y; t) \right>}{\delta h(\vec{x}\;',y';s)}
\right|_{h=0}
\nonumber
\end{eqnarray}
where we assumed spatial translation invariance in the directions parallel to the surface. For $y$, $y' \longrightarrow
\infty$ we recover the bulk quantities, whereas $y=y'=1$ yields the surface correlation and response functions. Of special
interest are the surface autocorrelation and autoresponse functions with $\vec{x}= \vec{x}\;'$ that we write as $C_1(t,s) :=
C(t,s;1,1,\vec{0}\;)$ and $R_1(t,s) := R(t,s;1,1,\vec{0}\;)$. For these surface quantities, the simple scaling forms
\begin{eqnarray}
\label{scaling_forms}
C_1(t,s) = s^{-b_1} f_{C_1}(t/s), &\quad & f_{C_1}(t/s)
\sim (t/s)^{-\lambda_{C_1}/z} \\ \nonumber
R_1(t,s)  =  s^{-1-a_1} f_{R_1}(t/s), &\quad &
f_{R_1}(t/s) \sim (t/s)^{-\lambda_{R_1}/z} 
\end{eqnarray}
are expected \cite{Pleimling04,Calabrese04} when $t,s$ and
also the difference $t-s$ are large compared to some
microscopic timescale. The scaling functions $f_{C_1}(t/s)$ and $f_{R_1}(t/s)$ should again display a simple power-law behavior for large
values of $t/s$.
General scaling arguments \cite{Calabrese04} allow to express the surface exponents appearing in (\ref{scaling_forms})
through other known exponents:
\begin{equation} \label{eq:exp_semi}
a_1  =  b_1 = (d - 2 +\eta_\parallel)/z ~~\mbox{and} ~~ \lambda_{C_1} = \lambda_{R_1} = \lambda_C + \eta_\parallel -\eta
\end{equation}
where $\eta_\parallel$ is the static exponent that governs the decay of the correlations parallel to the surface.
As for bulk systems \cite{Godr02,Cri03}, surface autocorrelation and autoresponse functions can be combined to yield
the surface fluctuation-dissipation ratio \cite{Pleimling04}
\begin{equation} \label{eq:x1}
X_1(t,s) = \frac{T_c R_1(t,s)}{\partial_s C_1(t,s)},
\end{equation}
with 
a universal limit value 
\begin{equation} \label{eq:x1infty}
X_1^{\infty} = \lim_{s \longrightarrow \infty} \left( \lim_{t \longrightarrow \infty} X_1(t,s) \right)
\end{equation}
that characterizes the different dynamical surface universality classes \cite{Calabrese04}.
The scaling picture (\ref{eq:c_r_semi}) and the relations between the various nonequilibrium exponents have
been verified by one of us through a numerical study of the out-of-equilibrium dynamics of various critical 
semi-infinite Ising models \cite{Pleimling04}.
In addition, the critical semi-infinite Gaussian model \cite{Calabrese04} and the critical
semi-infinite spherical model \cite{Baumann06} were also found to display
this simple aging scenario.

Whereas at least some knowledge has accumulated in recent years on the local aging behavior close to critical surfaces, almost nothing
is known on surface aging processes taking place in coarsening systems. In Ref. \cite{Baumann06} we have looked at the
out-of-equilibrium dynamical behavior of the semi-infinite spherical model.
For this special model
we have verified the existence of dynamical scaling and simple aging close to surfaces for quenches inside the ordered phase.
Surprisingly, the nonequilibrium exponent $b_1$, describing the scaling behavior of the
surface autocorrelation, was found to take on the value $b_1=1$, different from the standard value $b=0$
of the corresponding exponent in bulk systems undergoing phase-ordering. This result calls for a thorough investigation
of surface aging phenomena in other semi-infinite systems with phase-ordering dynamics.

In this paper we continue our study of local aging processes in bounded ferromagnets. On the one hand, we discuss 
the semi-infinite short-range Ising model in the limit of high dimensions
that can be solved exactly. On the other hand, we present 
results of extensive Monte Carlo simulations of the standard two-dimensional semi-infinite Ising model prepared at high
temperatures and then quenched inside the ordered phase. These numerical results yield new and interesting insights
into the local processes taking place in coarsening systems close to surfaces.
All the systems studied have in common that the dynamical exponent takes on the value $z=2$.

The paper is organized as follows. In the next Section we compute scaling functions and nonequilibrium
exponents in the exactly solvable semi-infinite model. Our numerical results obtained from simulations
of the two-dimensional
semi-infinite Ising model undergoing phase-ordering are then presented in Section III. Finally, in Section IV we draw
our conclusions and summarize our results.

\section{Quenching semi-infinite systems from high temperatures: exact results}
Exactly solvable models are often quite unrealistic and even artificial.
One of the reasons for nevertheless studying this kind of models in physics
is to obtain a guidance for the development of a future more sophisticated theoretical approach. 
With this in mind, we discuss in the following the nonequilibrium dynamical behavior of the exactly solvable
short-range semi-infinite Ising model in the limit of high dimensions that is prepared in an uncorrelated
initial state with vanishing magnetization and then quenched below or at the critical point. 
As in that limit the model is mean-field like, we expect the same critical exponents
and the same scaling functions (up to some numerical prefactors) as those found in the 
semi-infinite Gaussian model \cite{Calabrese04}.

The out-of-equilibrium behavior of the bulk Ising model
with nearest neighbor ferromagnetic interactions has recently been studied in the
limit of a large number $d$ of space dimensions \cite{Garriga05}.
Here we generalize the calculations of Garriga et al.
to the semi-infinite case. 

Using a semi-infinite hypercube with lattice constant 1, the Hamiltonian of our model can be written 
in the very general form
\begin{equation} \label{hamil_ising_d}
{\mathcal H} = - \frac{J_s}{2d} \sum_{(\vec{x},\vec{x}\;')} \sigma_{\vec{x},1} \sigma_{\vec{x}\;',1}
- \frac{J_b}{2d} \sum_{y \geq 2}\sum_{(\vec{x},\vec{x}\;')} \sigma_{\vec{x},y} \sigma_{\vec{x}\;',y}
- \frac{J_b}{2d} \sum_{y \geq 1} \sum_{\vec{x}} \sigma_{\vec{x},y} \sigma_{\vec{x},y+1} 
\end{equation}
where the sum over $(\vec{x},\vec{x}\;')$ indicates a sum over all nearest neighbor pairs lying in the
same layer. The spins can take on the values $\pm 1$, and an additional field term can be added if needed.
In writing (\ref{hamil_ising_d}) we take into account the layered structure of the lattice and distinguish
between nearest neighbor pairs lying in a layer parallel to the surface and nearest neighbor pairs
belonging to different layers.
As usual when dealing with semi-infinite systems \cite{Pleimling04}, we have introduced a different
coupling constant $J_s$ for interactions between nearest neighbor spins located both in the surface layer. 
We will however restrict ourselves in the following to the special case $J_s=J_b=1$. On the one hand this
yields in the limit $d \longrightarrow \infty$ the critical temperature $T_c = 1$
(where we set $k_B=1$), on the other hand we then
encounter at the critical temperature the so-called ordinary transition \cite{Pleimling04} where the
bulk alone is critical.

The main difference between the present case and the model considered in Ref.\cite{Garriga05} is of course
the absence of spatial translation invariance in the direction perpendicular to the surface. Due to this,
the time-dependent local fields that the spins experience are now layer-dependent, leading to the expressions
\begin{eqnarray} \label{eff_field}
h_{\vec{x},y}(t) & = & h^{ext}_{\vec{x},y}(t) +
\frac{1}{2d}\left(\sigma_{\vec{x},y+1}(t) 
+ \sum_{\vec{x}'(\vec{x})}\sigma_{\vec{x}',y}(t) \right) ~~\mbox{for} ~ y = 1,
\nonumber \\
h_{\vec{x},y}(t) & = & h^{ext}_{\vec{x},y}(t) +
\frac{1}{2d}\left(\sigma_{\vec{x},y+1}(t) +
\sigma_{\vec{x},y-1}(t)
+ \sum_{\vec{x}'(\vec{x})}\sigma_{\vec{x}',y}(t) \right) ~~\mbox{for} ~ y \neq 1,
\end{eqnarray}
where the sum over $\vec{x}\;'(\vec{x})$ is the sum over the in-plane nearest neighbor lattice sites
$\vec{x}\;'$ of $\vec{x}$.
We added in these equations an external field $h^{ext}_{\vec{x},y}(t)$ needed for the computation
of the response function (for the computation of the correlation function, $h^{ext}_{\vec{x},y}(t)$ is of course
set to zero).
Using heat-bath dynamics, these local fields $h_{\vec{x},y}(t)$ (which depend on the dimension $d$)
appear in the flip rates, as each spin
will flip independently with the rate  $( 1 - \sigma_{\vec{x},y}(t) \tanh(h_{\vec{x},y}(t)/T))/2$.
It is important to note that in absence of an external magnetic field the magnetization remains at any time at is initial value zero
everywhere in the sample.

\subsection{The correlation function}
In their paper \cite{Garriga05} Garriga et al. derived very general equations of motion for the one- and the two-time correlation
functions ${\mathcal C}$ and $C$ that can also be used in our case by plugging in the layer-dependent
local fields (\ref{eff_field}). Recalling that we still have invariance for spatial translations
parallel to the surface, we can write the following equations:
\BEA \label{eq_corr}
\partial_t {\mathcal C}(t;y,y',\vec{x}-\vec{x}\;') &=& - 2 {\mathcal C}(t;y,y',\vec{x} -\vec{x}\;')  +
\langle \Delta \mbox{t}_{\vec{x},y}(t) \Delta
\sigma_{\vec{x}\;',y'}(t) \rangle \nonumber \\ &+& \langle \Delta
\sigma_{\vec{x},y}(t) \Delta \mbox{t}_{\vec{x}\;',y'}(t)\rangle  \\
\partial_t C(t,s;y,y',\vec{x} - \vec{x}\;') &=& -
C(t,s;y,y',\vec{x} - \vec{x}\;')  + \langle \Delta \mbox{t}_{\vec{x},y}(t) \Delta
\sigma_{\vec{x}\;',y'}(s) \rangle 
\EEA
where we use the notations $\Delta \mbox{t}_{\vec{x},y}(t) := \tanh(h_{\vec{x},y}(t)/T) - \langle 
\tanh(h_{\vec{x},y}(t)/T) \rangle$ and $\Delta \sigma_{\vec{x},y}(t) = \sigma_{\vec{x},y}(t) - \langle
\sigma_{\vec{x},y}(t) \rangle$ for the deviations from the averages.

In the limit of large $d$ we can develop $\tanh(h_{\vec{x},y}(t)/T)$ in $1/d$ (recall that no external field is acting
on the spins and that the local fields $h_{\vec{x},y}$ are layer-dependent) which then yields
the following expressions for the equations of motion:
\pagebreak \typeout{********here is a pagebreak******}
\BEA
\label{diffeq_onetimecorr}
\partial_t {\mathcal C}(t;y,y',\vec{x}) &=& - 2 {\mathcal C}(t;y,y',\vec{x})  +
\frac{\gamma}{2}\Big({\mathcal C}(t;y+1,y',\vec{x}) + {\mathcal C}(t;y-1,y',\vec{x}) 
\\ & + & {\mathcal C}(t;y,y'+1,\vec{x}) + {\mathcal C}(t;y,y'-1,\vec{x})
+2 \sum_{\vec{z}(\vec{x})}{\mathcal C}(t;y,y',\vec{z})\Big) +
b(t;y,y',\vec{x}) \nonumber  \\
\label{diffeq_twotimecorr}
\partial_t C(t,s;y,y',\vec{x}) &=& -C(t,s;y,y',\vec{x})  +
\frac{\gamma}{2}\Big(C(t,s;y+1,y',\vec{x}) \nonumber \\  
& + &C(t,s;y-1,y',\vec{x})+ \sum_{\vec{z}(\vec{x})}C(t,s;y,y',\vec{z})\Big).
\EEA
with
\BEQ
\label{boundary_cond}
C(t,s;0,y',\vec{x}) = 0 = C(t,s;y,0,\vec{x})~.
\EEQ
In writing these equations we exploit the spatial translation invariance parallel to the surface by setting 
$\vec{x}\;'=\vec{0}$. The parameter $\gamma$ is given by $\gamma := 1/(T\,d)$, whereas the sum
over $\vec{z}(\vec{x})$ indicates a summation over the in-plane nearest neighbor lattice sites of $\vec{x}$. 
The quantity $b(t;y,y',\vec{x}) = \delta_{y,y'}
\delta_{\vec{x},\vec{0}} \, \bar{b}(t;y)$, which is needed to enforce
the condition ${\mathcal C}(t;y,y,\vec{0}) = 1$
for all times $t$, has to be determined self-consistently \cite{Garriga05}. In addition, 
the two-time correlator must yield the one-time
correlator for $t = s$, i.e.\
\BEQ
C(t,t;y,y',\vec{x}) = {\mathcal C}(t;y,y',\vec{x}).
\EEQ
The solution of these equations of motion is outlined in the Appendix.
For decorrelated initial conditions, our result is:
\BEA
\label{correlator_exact}
C(t,s;y,y',\vec{x}) &=& e^{-(t+s)} \Big(I_{y-y'}(\gamma(t+s)) -
I_{y+y'}(\gamma(t+s))\Big) \prod_{i=1}^{d-1}
I_{x_i}(\gamma(t+s))\nonumber \\
&+& \sum_{u \geq 1} \int_0^s d\,\tau\,\, \bar{b}(\tau,u)\, e^{-(t+s-2\tau)}
\prod_{i=1}^{d-1} I_{x_i}(\gamma(t+s -2 \tau))
\nonumber \\ & & \hspace{-2.0cm} \times 
\Big(I_{u-y}(\gamma(t -\tau)) -
I_{u+y}(\gamma(t-\tau)) \Big)
\Big(I_{u-y'}(\gamma(s-\tau)) - I_{u+y'}(\gamma(s-\tau)) \Big)
\EEA
where the functions $I_\nu$ are modified
Bessel functions \cite{Gradshteyn80} and 
where we have taken into account the special form of
$b(t;y,y',\vec{x})$ and Equation (\ref{bessel_identity}). 
It remains to fix the parameter
$\bar{b}(t,y)$, which we determine from the
condition ${\mathcal C}(t;y,y;\vec{0}) = 1$. For large $d$ the factor
$\gamma = 1/(Td)$ becomes small, and we can use the following approximation
\BEQ
I_{y-y'}(\gamma (t-\tau)) \approx \delta_{y,y'} + O
\left(\frac{1}{d}\right)
\EEQ
and similarly for other terms, see also Ref. \cite{Garriga05}.
This yields for vanishing layer magnetization the equation
\BEQ
1 = e^{-2t} + \int_0^td\,\tau \,e^{-2(t-\tau)} \bar{b}(\tau,y)
\EEQ
for all $y$ and $t$. This equation can be
solved by Laplace transform, yielding the result
$\bar{b}(t,y) = 2$ for all $y$ and $t$.
It then follows that the correlation function in
the semi-infinite model is given by Eq. (\ref{correlator_exact}) with
$\bar{b}(t,u)$ set to $2$. One can get rid of the
sum over $u$ by using $\sum_{m=-\infty}^\infty I_{m+k}(z_1)
I_m(z_2) = I_k(z_1 + z_2)$ and $I_n(z) = I_{-n}(z)$. 
After doing so, we obtain for the surface
autocorrelation function the expression
\BEA \label{surf_corr_final}
& &C_1(t,s) = e^{-(t+s)} \Big(I_0(\gamma(t+s))\Big)^{d-1}
\Big(I_0(\gamma(t+s)) - I_{2}(\gamma(t+s))\Big) \nonumber
\\& & + 2 \int_0^s d\,\tau \, e^{-(t+s-2 \tau)} \Big(I_0(\gamma(t+s-2
\tau))\Big)^{d-1}\Big(I_0(\gamma(t+s -\tau)) - I_2(\gamma(t+s-2
\tau))\Big).
\EEA
It is worth noting that in the limit where $y \approx y' \rightarrow \infty$
we also recover the known bulk behavior of the
correlation function, as we get with the help of expression
(\ref{bessel_identity}) the expression
\BEA
C(t,s;y,y',\vec{x}) &=& e^{-(t+s)} I_{y-y'}(\gamma(t+s)) \prod_{i=1}^{d-1}
I_{x_i}(\gamma(t+s))\nonumber \\
&+& 2  \int_0^s d\,\tau\,\, e^{-(t+s-2\tau)}
I_{y-y'}(\gamma(t+s-2 \tau))
\prod_{i=1}^{d-1} I_{x_i}(\gamma(t+s -2 \tau)) 
\EEA
which is precisely the expression found by Garriga
et al. \cite{Garriga05} in Fourier-space.

We immediately remark that for a quench inside the ordered phase with
$T < T_c=1$ no simple scaling behavior of the surface autocorrelation 
is observed, due to the extremely
rapidly (i.e. exponentially) increasing Bessel functions. A similar absence
of dynamical scaling is also seen in the bulk system quenched below
the critical point \cite{Garriga05}.
At the critical point
however, when $T=1$ and therefore $\gamma = 1/d$, we can use the approximation 
$e^{-u} I_\nu(u) \approx (2 \pi u)^{-1/2} \exp(-\nu^2/(2 u))$. 
As the first term in (\ref{surf_corr_final}) decreases more rapidly than the second one, we find
in the scaling regime (with $Y = t/s$)
\BEQ
\label{surf_autocorr}
C_1(t,s) = 4 \left(\frac{2 \pi}{d}\right)^{-d/2} s^{d/2}
\Big((Y-1)^{-d/2}-(Y+1)^{-d/2} \Big).
\EEQ
This allows us to identify both the nonequilibrium exponents $b_1$ and $\lambda_{C_1}$ and
the scaling function $f_{C_1}(Y)$, see Eq. (\ref{scaling_forms}):
\BEQ
b_1 = \frac{d}{2}, \qquad \lambda_{C_1} = d + 2
,\qquad f_{C_1}(Y) =  4 \left(\frac{2
\pi}{d}\right)^{-d/2}\Big( (Y-1)^{-d/2} - (Y+1)^{-d/2}
\Big)
\EEQ
where we used that in the limit of large $d$ the critical dynamical exponent is equal to 2.

\subsection{The response function}

In order to compute the response function we start from the
differential equation \cite{Garriga05}
\BEQ
\partial_t \langle
\sigma_{\vec{x},y}(t) \rangle = - \langle
\sigma_{\vec{x},y}(t) \rangle + \langle  \mbox{t}_{\vec{x},y}(t) \rangle
\EEQ
for $\langle
\sigma_{\vec{x},y}(t) \rangle$ in the presence of a small
external magnetic field $h^{ext}_{\vec{x},y}$.
As both $h^{ext}_{\vec{x},y}$ and $1/d$ are small we can develop the $\tanh$ to
first order in both quantities:
\BEQ
\tanh(h_{\vec{x},y}(t)/T) \approx \frac{1}{T} h^{ext}_{\vec{x},y}(t)
+ \frac{\gamma}{2} \Big( \sigma_{\vec{x},y+1}(t)
+ \sigma_{\vec{x},y-1}(t) +  \sum_{\vec{z}(\vec{x})}
\sigma_{\vec{z},y}(t) \Big).
\EEQ
where it is understood that $\sigma_{\vec{x},0} = 0$. The definition
\BEQ
R(t,s;y,y';\vec{x} - \vec{x}\;') := \frac{\delta \langle
\sigma_{\vec{x},y}(t)\rangle}{\delta h^{ext}_{\vec{x}\;',y'}(s)}
\EEQ
of the response function now directly yields the
differential equation (where we set again $\vec{x}\;' = \vec{0}$)
\BEA
\partial_t R(t,s;y,y',\vec{x}) &=& - R(t,s;y,y',\vec{x}) +
\frac{\gamma}{2} \Big(R(t,s;y+1,y',\vec{x}) +
R(t,s;y-1,y',\vec{x}) \nonumber \\ & + & \sum_{\vec{z}(\vec{x})}
R(t,s;y,y',\vec{z}) \Big) + \frac{1}{T} \delta(t-s)
\delta_{y,y'} \delta_{\vec{x},\vec{0}}
\EEA
with $R(t,s;0,y',\vec{x}) = 0 = R(t,s;y,0,\vec{x})$.
This equation is solved with similar methods as outlined in
the Appendix for the correlation function. As a result we obtain
\BEQ
R(t,s;y,y',\vec{x}) = \frac{\Theta(t-s)
}{T}e^{-(t-s)} \Big(I_{y-y'}(\gamma(t-s)) - I_{y+y'}(\gamma(t-s))\Big)
\prod_{i=1}^{d-1} I_{x_i}(\gamma(t-s))
\EEQ
For the case $T = T_c=1$ the surface autoresponse function
can again be evaluated in the scaling regime, yielding
(with $Y = t/s$)
\BEQ
\label{surf_autoresp}
R_1(t,s) = \frac{2d}{T_c} \left(\frac{2
\pi}{d}\right)^{-d/2}s^{-\frac{d}{2}-1}(Y-1)^{-(d+2)/2}
\EEQ
and therefore
\BEQ
a_1 = \frac{d}{2}, \qquad \lambda_{R_1} = d + 2, \qquad
f_{R_1}(Y) = \frac{2d}{T_c} \left(\frac{2
\pi}{d}\right)^{-d/2}(Y-1)^{-\frac{d}{2}-1}.
\EEQ
We can now also compute the surface fluctuation-dissipation ratio
from the expressions (\ref{surf_autocorr}) and (\ref{surf_autoresp}) and obtain
\BEA
X_1(t,s) &=& \frac{T_c R_1(t,s)}{\partial_s C_1(t,s)}
\nonumber \\  & = &
 \frac{(Y-1)^{-\frac{d}{2}-1}}{Y \Big((Y-1)^{-\frac{d}{2}-1} -
 (Y+1)^{-\frac{d}{2}-1}\Big) - \Big((Y-1)^{-\frac{d}{2}} -
 (Y+1)^{-\frac{d}{2}}\Big)}
\EEA
from which the limit value $X_1^\infty = 1/2$ follows.

Comparing with the results obtained for the spherical model, see Table \ref{table1}, we note
that the values of the nonequilibrium exponents in the critical short-range Ising model in the limit of high dimensions are
in full agreement with the values obtained for the critical spherical model in dimensions $d>4$ \cite{Baumann06}. Even the
scaling functions are identical up to a nonuniversal numerical prefactor.
Besides, the values of the universal quantities are identical
to the values obtained in the field-theoretical Gaussian 
model \cite{Calabrese04}. 
This nicely demonstrates that in the aging regime universal nonequilibrium features are indeed
encountered close to critical surfaces.

\section{Quenching semi-infinite systems from high temperatures: numerical results}
In the following we present the results of extensive numerical simulations of the standard 
two-dimensional semi-infinite Ising model
with only nearest-neighbor interactions quenched inside the ordered phase. 
In the bulk case this model is known to render rather faithfully the physics of real systems undergoing phase
ordering.
The Hamiltonian is given by the usual expression
\begin{equation}
{\mathcal H} = - J \sum\limits_{\langle i, j \rangle} \sigma_i \sigma_j~,
\end{equation}
where $i$ and $j$ label the sites of a semi-infinite lattice. The sum extends over nearest neighbor pairs, and
we have the same coupling strength $J>0$ for every bond connecting neighboring spins. This system exhibits a continuous
phase transition at the bulk critical point $T_c = 2/\ln(\sqrt{2} +1 ) \approx 2.269$ (where the temperature is measured
in units of $J/k_B$, with $k_B$ being the Boltzmann constant).

Whereas surface aging behavior has already been studied in the past for Ising models quenched onto the critical point
\cite{Pleimling04}, this does not seem to be the case for quenches below the critical point. The following numerical
study therefore allows us to close a gap in our understanding of the nonequilibrium dynamical behavior of classical
spin models. Especially, it yields new insights into the local dynamical behavior of systems undergoing phase-ordering
in the presence of surfaces.

For these simulations we use periodic boundary conditions in one direction and free boundary conditions in the other
direction. We thereby consider square systems with $N=L \times L$ spins were $L$ ranges from $L=300$ to $L=1000$, thus
making sure that the data obtained at any one of the two surfaces are representative of the semi-infinite system. Only data 
free of finite-size effects are discussed in the following. Our focus lies on the surface autocorrelation function
and on the surface autoresponse function. The surface autocorrelation function is
given by the expression
\begin{equation}
C_1(t,s) = \frac{1}{2L}\sum\limits_{i \in surface} \langle \sigma_i(t) \sigma_i(s) \rangle,
\end{equation}
where the sum is over all the spins in the two surfaces. The data discussed in the following have been obtained 
after averaging over at least 5000 different runs with different realizations of the noise.
In order to study the response to a magnetic field, we apply
a weak binary random field between the time $t=0$
(at which the quench takes place) and the time $t=s$ \cite{Bar98}. After the field has been switched off, we monitor the decay of
the surface thermoremanent magnetization given by the expression
\begin{equation}
M_1(t,s)= \frac{1}{2L} \, \sum\limits_{i \in surface} \overline{\langle
h_i \, \sigma_i(t) \rangle}/T,
\end{equation}
where $h_i$ is the strength of the binary random field at site $i$. In addition to averaging over the 
realizations of the noise we also average
over the realizations of the random field as indicated by the bar. We discuss here data obtained with $|h_i|=0.1$ (we checked that 
our conclusions remain the same when we slightly vary the value of $|h_i|$). As response functions are very noisy, we 
average over many more runs than for the autocorrelation. The thermoremanent magnetization data
discussed in this section have been obtained after averaging over typically 200,000 runs.

\subsection{Autocorrelation function}
Before discussing the surface autocorrelation function, let us briefly mention some results obtained for the autocorrelation
function in the corresponding two-dimensional bulk system. The expected scaling form
\begin{equation}
C(t,s) = s^{-b} f_C(t/s) ~~\mbox{with} ~~ f_C(t/s) \sim (t/s)^{-\lambda_C/z} ~~ \mbox{for} ~~ t/s \gg 1
\end{equation}
has been verified in various numerical studies. These studies showed that $b=0$ and yielded the value $\lambda_C/z = 0.63(1)$
\cite{Fish88,Brow97,Henkel03}
(recall that $z=2$) for the exponent governing the long-time decay of the scaling function.
Numerous theoretical approaches have been proposed for computing the scaling function $f_C$ 
\cite{Bray91a,Bray92,Liu91,Roja99,Maze98}, the most successful being the
recent exploitation of space-time symmetries within the theory of local scale invariance \cite{Henkel04,Lorenz07,HenBau07}.

The main question we address here concerns the scaling behavior of the surface autocorrelation function. Let us
start by looking at the long-time decay of $C_1(t,s)$ with $s=0$, as it is well known that this quantity is usually
the most appropriate for the determination of $\lambda_{C_1}$. In Figure 1 we show this quantity for two 
different temperatures, $T=1$ and $T=1.5$, lower than the critical temperature. For comparison we also include
the bulk autocorrelation function $C(t,s=0)$ for the same two temperatures. Whereas at short times the surface autocorrelation
(this is also true for the bulk quantity) is clearly temperature dependent, at longer times the two curves 
get identical. Interestingly, the decay of the surface correlations follow a power-law at late times.
This power-law decay is faster at the surface than inside the
bulk, yielding the value $\lambda_{C_1}/z = 0.95(3)$ which should be compared to the value $\lambda_C/z = 0.63(1)$ 
obtained inside the bulk. Obviously, this faster decay is due to the reduced coordination number at the surface.

\begin{figure}[t] \label{fig1}
\centerline{\psfig{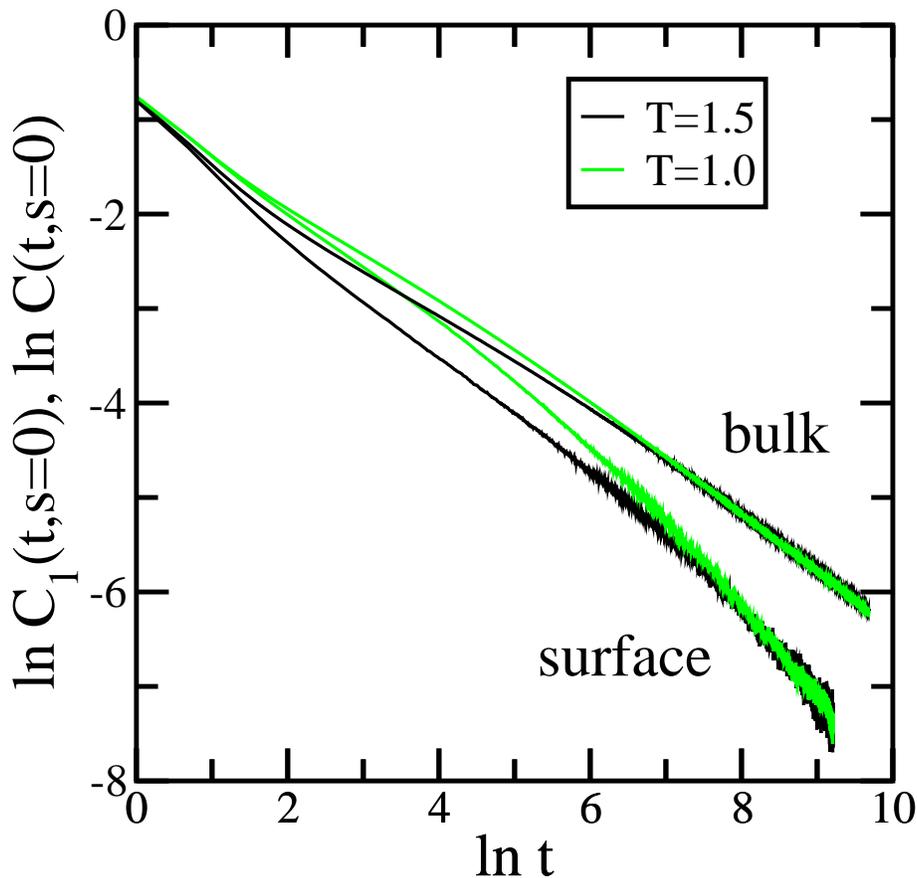}}
\caption{(Color online) 
The surface and bulk autocorrelation functions $C_1(t,s=0)$ and $C(t,s=0)$ for two different temperatures. At the surface the
correlations decay much faster than inside the bulk, yielding the value $\lambda_{C_1}/z = 0.95(3)$ for the long-time
power-law exponent, considerably larger than the value $\lambda_C/z = 0.63(1)$ in the bulk case.
}
\end{figure}

In Figure 2 we discuss the behavior of the surface autocorrelation function $C_1(t,s)$ with $s>0$.
When plotting $C_1(t,s)$ versus $t/s$, we do not observe a data collapse, see Figure 2a,
in contrast to the data collapse observed when plotting the bulk autocorrelation
as a function of $t/s$. The data shown in Figure 2a at first look suggest that
the local exponent $b_1$ is different from zero at the surface. A more thorough analysis reveals however that 
a good scaling behavior can not be achieved with a constant $b_1 > 0$. Figure 2b shows our best result
obtained for $b_1 =0.13$. A reasonable data collapse can be achieved this way for large values of $t/s$, but
scaling breaks down for $t/s \leq 25$. Taken at face value, this would suggest for the surface autocorrelation function
the existence of a large threshold value
of $t/s$ below which dynamical scaling is not observed. The possible physical
mechanism responsible for this threshold is far from obvious. A better data collapse can be achieved by allowing the
exponent $b_1$ to depend itself on $t/s$, but a non-constant exponent varying as a function of $t/s$ is not supported 
by any theoretical approach.

\begin{figure}[!h] \label{fig2}
\centerline{\psfig{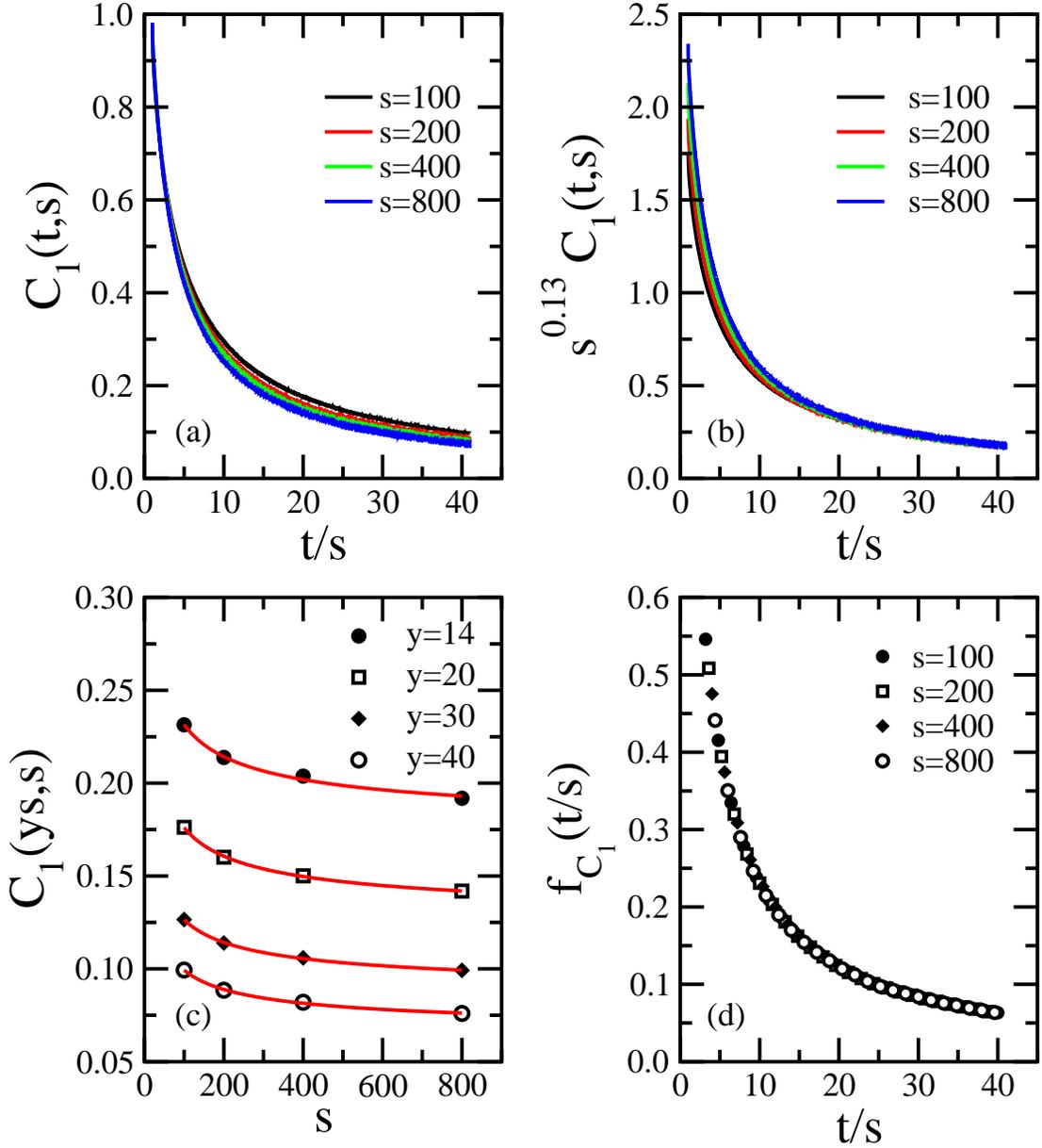}}
\caption{(Color online) 
Discussion of the surface autocorrelation $C_1(t,s)$ obtained after quenching the semi-infinite two-dimensional
Ising model to $T=1$. (a) Autocorrelation as a function of $t/s$ for different values of the waiting time $s$ 
(the lowest curve shows the data obtained for the largest value of $s$). The expected data collapse with $b_1 =0$ is not
observed. (b) Plotting $s^{0.13} C_1(t,s)$ versus $t/s$ leads to a collapse of data for large values of
$t/s$, but no scaling is observed for smaller values of $t/s$. (c) Plot of $C_1(ys,s)$ as a function of $s$ for
various values of $y=t/s$. The full lines are fits to the extended scaling form (\ref{eq:finite_time}) with $b' = 0.49$.
(d) Scaling function $f_{C_1}(t/s)$ obtained from the data shown in (a) after subtracting off the
finite-time correction term. In (c) and (d) error bars are smaller than the symbol sizes.
}
\end{figure}

We propose here another interpretation of the numerical data that
is based on the recent observation
that large finite-time corrections can to some extend mask the true scaling behavior of the autocorrelation function in
phase-ordering systems \cite{Hen07}. In order to take the existence of finite-time corrections into account, we try to
describe our data by the ansatz
\begin{equation} \label{eq:finite_time}
C_1(t,s) = f_{C_1}(t/s) + s^{-b'} g_{C_1}(t/s)~,
\end{equation}
where the first term is the expected scaling behavior with $b_1 =0$, whereas the second term is the finite-time correction
that is of decreasing importance for increasing values of the waiting time $s$. This ansatz has recently been used for the
analysis of the autocorrelation functions in disordered ferromagnets quenched below their critical point \cite{Hen07,Henkel06}.
In Figure 2c we show $C_1(ys,s)$ as a function of $s$ for various values of the ratio $y=t/s$. 
The lines show that an excellent fitting of the data can be achieved with the extended scaling form (\ref{eq:finite_time})
with a common value $b'=0.49(1)$. The scaling function $f_{C_1}(t/s)$, obtained after subtracting off the
correction term, is shown in Figure 2d. As the curves for the different values of $s$ are not distinguishable on the
scale of the Figure, we only show selected points as symbols. The data collapse shown in Figure 2d supports our
interpretation that the true scaling behavior of the surface autocorrelation function is masked by strong
finite-time corrections. As a consistency check, we note that the data in Figure 2d present for large values
of $t/s$ a power-law decay with an exponent 0.95(2), in full agreement with the value of $\lambda_{C_1}/z$ obtained
directly from $C_1(t,s=0)$. Even though our data are perfectly described by Eq. (\ref{eq:finite_time}),
we must emphasize that we do not yet know why this finite-time 
correction shows up close to the surface but is not encountered inside the bulk.

Let us end the discussion of the surface autocorrelation function by noticing that the value $b' =0.49(1)$ 
of the correction term exponent is compatible
with $1/2=1/z$. However, we refrain from making the conjecture $b'=1/z$ here without having studied other systems with surfaces
(as for example semi-infinite Potts models).

\subsection{Response function}
Before discussing the surface thermoremanent magnetization $M_1(t,s)$, let us again first recall the behavior of the 
corresponding bulk quantity. The bulk thermoremanent magnetization $M(t,s)$ is a temporally integrated response 
function that is related to the response function $R(t,s)$ by the integral
\begin{equation}
M(t,s) = \int\limits_0^s du\, R(t,u)~,
\end{equation}
where the integration is over the whole time interval during which the magnetic field was acting on the system.
{}From the scaling form (\ref{eq_dyn_scal}) of $R(t,s)$, we therefore obtain the scaling behavior
\begin{equation}
M(t,s) = s^{-a} f_M(t/s)
\end{equation}
for the integrated response. Zippold, K{\"u}hn, and Horner \cite{Zippold2000} were the first to point out the existence of a subleading
correction term which can be quite sizeable.
For the thermoremanent magnetization this leads to the 
following more complete scaling behavior \cite{Henk02a},
\begin{equation}
M(t,s) = s^{-a} f_M(t/s) + s^{-\lambda_R/z} g_M(t/s)~.
\end{equation}
The second term in this equation is in fact the response of the system to fluctuations in the 
initial state, where the scaling function $g_M(t/s)$ is expected to be proportional to the power-law $(t/s)^{-\lambda_R/z}$
\cite{Humayun1991}. For the two-dimensional Ising model we have $a=1/z=1/2$ and $\lambda_R/z =0.63$. Therefore this
correction to scaling can not be neglected but must be included in order to obtain the correct description of
the scaling behavior of the bulk thermoremanent magnetization \cite{Henk02a,Lorenz07}.

In Figure 3 we summarize our findings for the surface thermoremanent magnetization in the two-dimensional
semi-infinite Ising model quenched below the critical point. Figure 3a shows the behavior of this local response
as a function of $t/s$ for various values of the waiting time $s$. In a first attempt, we might try to achieve
a scaling behavior by assuming that 
\begin{equation}
M_1(t,s) = s^{-a_1} f_{M_1}(t/s)~,
\end{equation}
thereby neglecting any possible corrections to scaling. A reasonable scaling behavior is achieved this way for
a value of $a_1 \approx 0.40$, slightly lower than the expected value
$1/z=1/2$. For a more thorough analysis we can fix $y=t/s$ and plot the response as a function of
the waiting time in a log-log-plot. Fitting a straight line to the data, we obtain from the slope of that line 
a value of $a_1$ 
for every considered value of $t/s$. Thus, we obtain $a_1 =0.38$ for $t/s = 5$, $a_1 = 0.39$ for $t/s=10$, 
$a_1 = 0.40$ for $t/s=15$, and $a_1 = 0.42$ for $t/s=20$. This points to the existence of a correction term that vanishes
for increasing values of $t/s$. 
In Figure 3b we test the more complete scaling form
\begin{equation} \label{ext_scal_m}
M_1(t,s) = s^{-a_1} f_{M_1}(t/s) + s^{-\lambda_{R_1}/z} g_{M_1}(t/s)~
\end{equation}
where the correction term with the scaling function $g_{M_1}(t/s) = r_1 (t/s)^{-\lambda_{R_1}/z}$ describes the response
of the surface to fluctuations in the initial state. Plugging in the value $\lambda_{R_1}/z = 0.95$ 
(where we assume that $\lambda_{R_1} = \lambda_{C_1}$ holds), we obtain a consistent description for any $t/s$
with common values $r_1 = -0.106(1)$ for the amplitude of the correction term and $a_1=0.50(1)$ for the exponent of
the leading term. The correction term being now completely fixed, we can subtract it off from the numerical data and 
obtain the data collapse shown in Figure 3c. Thus, as for the thermoremanent magnetization in the bulk
\cite{Henk02a,Lorenz07}, we are
able to identify the leading correction term and in addition obtain the value $a_1=a=1/z$.

\begin{figure}[!ht] \label{fig3}
\centerline{\psfig{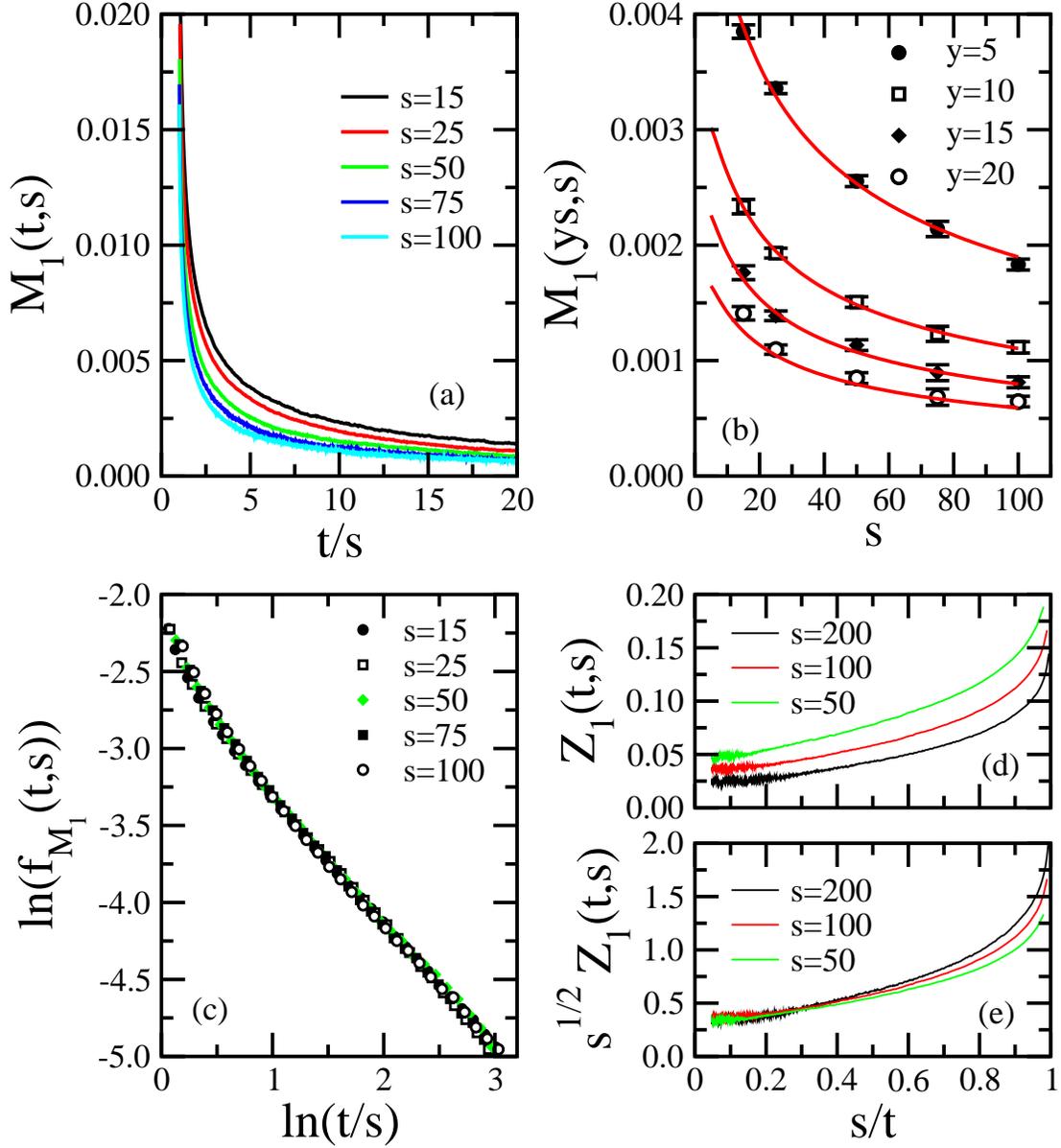}}
\caption{(Color online)
Discussion of the surface thermoremanent magnetization $M_1(t,s)$ obtained after quenching the semi-infinite two-dimensional
Ising model to $T=1$ where a random magnetic field of strength $h=0.1$ is applied between $t=0$ and 
$t=s$. (a) $M_1(t,s)$ plotted against $t/s$ for various waiting times $s$
(the lowest curve shows the data obtained for the largest value of $s$).
(b) Plot of $M_1(ys,s)$ as a function of $s$ for
various values of $y=t/s$. The full lines are fits to the extended scaling form (\ref{ext_scal_m}).
A consistent description of the
data for any value of $t/s$ is achieved for $\lambda_R/z = 0.95$, $a=0.5$, and $r_1 = -0.106$, see main text.
(c) Scaling function $f_{M_1}(t/s)$ obtained from the data shown in (a) after subtracting off the
finite-time correction term. Error bars are smaller than the symbol
sizes. (d) Plot of $Z_1(t,s)$, see Equation (\ref{eq:Z}), versus $s/t$ for different
waiting times. (e) Plotting $s^{1/2} Z_1(t,s)$ versus $s/t$ leads to a data collapse for small values of $s/t$.
}
\end{figure}

We close this section by a brief discussion of the surface fluctuation-dissipation ratio. In Figure 3d
we plot the ratio 
\begin{equation} \label{eq:Z}
Z_1(t,s) = \frac{T M_1(t,s)}{h C_1(t,s)}
\end{equation}
as a function of $s/t$ for various values of $s$.
This ratio yields asymptotically the limit value $X_1^\infty$ of the fluctuation-dissipation ratio (\ref{eq:x1}), as
\begin{equation}
X_1^\infty = \lim_{s \longrightarrow \infty} \left( \lim_{t \longrightarrow \infty} Z_1(t,s) \right)~.
\end{equation}
For a given value of the waiting time, the ratio $Z_1(t,s)$ converges towards a constant finite value when $s/t \longrightarrow 0$. 
At first look this might seem surprising as in coarsening systems one expects the limit value $X_1^\infty=0$. However, this
constant decreases for increasing values of $s$. Taking into consideration the leading scaling behaviors of $C_1(t,s)\sim
f_{C_1}(t/s)$ and of
$M_1(t,s) \sim s^{-1/2} f_{M_1}(t/s)$ found in our study as well as the fact that the scaling functions
$f_{C_1}(t/s)$ and $f_{M_1}(t/s)$ display for large arguments a power-law behavior with the same
exponent $0.95$, we find that the saturation value $\lim_{t \longrightarrow \infty}
Z_1(t,s)$ should vanish as $s^{-1/2}$. This is indeed verified in Figure 3e, where $s^{1/2} Z_1(t,s)$ leads to a collapse 
of the data onto a common curve  for $s/t$ small. This is also an {\it a posteriori} check that we have indeed correctly
identified the leading scaling behaviors of both the surface autocorrelation and the surface integrated response functions.

\section{Conclusions}
In this paper we have extended the investigation of surface aging phenomena to cases not 
studied previously. On the one hand we have computed nonequilibrium surface quantities 
in the exactly solvable short-range semi-infinite Ising model in the limit of a large number
of space dimensions, 
on the other hand we have presented numerical simulations of the standard semi-infinite Ising
model quenched inside the ordered phase. 

For a quench to the critical point, we added the semi-infinite Ising model in high dimensions to the list
of exactly solved models. The universal nonequilibrium surface quantities obtained in this study agree with 
those obtained for the critical semi-infinite spherical model \cite{Baumann06}, as expected for a 
mean-field like model. In Table I we summarize the known results for surface aging phenomena in critical systems
at the ordinary transition (the only situation studied in this paper)
by listing the values of the different universal nonequilibrium exponents as well as those of the asymptotic
value of the fluctuation-dissipation ratio. It is worth mentioning that the existing numerical data
for the semi-infinite Ising model \cite{Pleimling04}
indicate a non-monotonic behavior of the limit value of the surface fluctuation-dissipation ratio
as a function of the dimensionality of the system (being $\frac{1}{2}$ for $d \geq 4$, then increasing to $0.59$ in three
dimensions, before decreasing to $0.31$ in the two-dimensional system). This behavior is unexpected, and a satisfactory
explanation is still lacking.

\begin{table}
\caption{\label{table1} Available values of nonequilibrium critical surface quantities 
at the ordinary transition determined in aging systems
quenched to the critical point.
}
\begin{displaymath}
\begin{array}{||c||c|c|c||} \hline \hline
\mbox{model} & 
 a_1 = b_1  & \lambda_{R_1} = \lambda_{C_1} &
X_1^\infty \\ \hline \hline
\mbox{spherical model ($ 2 < d <4$) } \cite{Baumann06} & \frac{d}{2} & \frac{3
d}{2}  & 1 - \frac{2}{d} \\
\mbox{spherical model ($d > 4$)} \cite{Baumann06} & \frac{d}{2} & d + 2 & \frac{1}{2}  \\
\mbox{Gaussian model} \cite{Calabrese04} & \frac{d}{2} & d + 2 & \frac{1}{2} \\
\mbox{Ising model in large dimensions} & \frac{d}{2} & d + 2 & \frac{1}{2}  \\
\mbox{Ising model in $d=3$, ordinary transition} \cite{Pleimling04} & 1.24(1)  & 2.10(1)  & 0.59(2)  \\
\mbox{Ising model in $d=2$} \cite{Pleimling04} & 0.46(1)  & 1.09(1)  & 0.31(1)  \\ 
\hline \hline
\end{array}
\end{displaymath}
\end{table}

We also presented large-scale numerical simulations of the two-dimensional semi-infinite Ising model 
undergoing coarsening. From these results we conclude that the result $b_1 \neq 0$ 
found for the spherical model \cite{Baumann06}
is not generic but that it is very probably an artifact of that rather artificial model.
Indeed, the numerical simulations of the more realistic two-dimensional Ising model yield $b_1 = 0$. This indicates
that generically the exponent $b_1$, that governs the scaling of the surface correlations, vanishes, similarly to what
is observed inside the bulk.

One of the main conclusions of our work is that surface aging phenomena in systems undergoing phase-ordering display
the same general features as bulk aging phenomena. Simple scaling forms prevail asymptotically for two-time quantities like
the surface autoresponse and the surface autocorrelation functions, and universal nonequilibrium quantities,
with values that differ
from the bulk values,
can also be identified in semi-infinite coarsening systems, see Table 2. 
For finite times, corrections to scaling can be rather important and
might even mask the leading scaling behavior. In our study of the two-dimensional semi-infinite 
Ising model we not only identified a sub-leading contribution to the thermoremanent surface magnetization (a similar
correction also appears inside the bulk), but we also showed the existence of corrections to scaling in the
surface autocorrelation function. The physical origin of this last term is not yet clear. It is however worth noting that a
similar correction term has recently been shown to exist for the random bond Ising model quenched below the
critical point \cite{Hen07}.

\begin{table}
\caption{\label{table2} Available values of nonequilibrium surface quantities determined in aging systems
quenched below the critical point.}
\begin{displaymath}
\begin{array}{||c||c|c|c||} \hline \hline
\mbox{model} & 
 a_1 & b_1  & \lambda_{R_1} = \lambda_{C_1} \\ \hline \hline
\mbox{spherical model} \cite{Baumann06} & \frac{d}{2} & 1 & \frac{d}{2}+2 \\
\mbox{Ising model in $d=2$} & \frac{1}{2}& 0 & 1.90(6) \\ \hline \hline
\end{array}
\end{displaymath}
\end{table}

The semi-infinite geometry discussed in this paper is of course only a special case of a more general wedge-shaped geometry.
Wedges in critical systems have been studied quite intensively in the past \cite{Car83,Igl93,Ple98,Ple04a}, 
as they lead to static critical quantities
whose values depend on the opening angle of the wedge. However, the local critical dynamical behavior in a wedge-shaped geometry
has not yet been discussed in the literature. Phase-ordering in wedges can also be viewed as being one of the simplest cases of
phase-ordering in confined geometries. The study of edge aging phenomena is therefore the next logical
step in the study of local nonequilibrium dynamical behavior in confined geometries, and work along this line is in progress.

\begin{acknowledgments}
We acknowledge the support by the Deutsche Forschungsgemeinschaft
through grant no. PL 323/2
and by the franco-german binational
programme PROCOPE. The numerical work was done on Virginia Tech's System X.
\end{acknowledgments}

\appsection{A}{Ising model in high dimensions: Computation of the correlation function}
In this Appendix we compute the one-time and two-times correlation functions for the semi-infinite
Ising model in high dimensions.
We thereby start by defining the operator $
\square_{t;y,y',\vec{x}}^{(\gamma)}$ for a function $f:
\mathbb{R} \times \mathbb{Z}_{\geq 0} \times \mathbb{Z}_{\geq 0} \times
\mathbb{Z}^{d-1} \rightarrow \mathbb{R} $:
\BEA
\square_{t;y,y',\vec{x}}^{(\gamma)} 
f(t;y,y',\vec{x}) &:=& \partial_t f(t;y,y',\vec{x}) + 2
f(t;y,y',\vec{x})  -\frac{\gamma}{2}\Big(f(t;y+1,y',\vec{x}) +
f(t;y-1,y',\vec{x}) \nonumber \\ &+ &f(t;y,y'+1,\vec{x}) + f(t;y,y'-1,\vec{x})
+2 \sum_{\vec{x}'(\vec{x})}f(t;y,y',\vec{x}')\Big)~,
\EEA
where $\vec{x}'(\vec{x})$ denotes the nearest neighbors of
$\vec{x}$ in the layer $y$. With this, Equation (\ref{diffeq_onetimecorr}) reads
\BEQ
\square_{t;y,y',\vec{x}}^{(\gamma)} \, C(t;y,y',\vec{x}) = b(t;y,y',\vec{x}).
\EEQ
In order to solve this equation we look for the Green's function
satisfying the equation
\BEQ
\label{diffeqn_gf}
\square^{(\gamma)}_{t;y,y',\vec{x}} g(t;u,y,v,y',\vec{x}) = \delta(t) \delta_{y,u}
\delta_{y',v} \delta_{\vec{x},0} 
\EEQ
and the boundary conditions
\BEQ
g(t;u,0,v,y',\vec{x}) = 0 = g(t;u,y,v,0,\vec{x}).
\EEQ
We can solve this equation by using a Fourier-Sine
transformation on $y$ and $y'$ (which makes sure the boundary conditions hold) 
and a normal Fourier transformation on $\vec{x}$:
\BEA
\label{f_trafo1}
\hat{g}(t;u,k,v,k',\vec{q}) & = & \sum_{y,y' \geq 0}
\sum_{\vec{x}} \sin(k\, y) \sin(k'\,y') e^{\II \vec{x} \cdot
\vec{q}} g(t;u,y,v,y',\vec{x})~, \\
\label{f_trafo2}
g(t;u,y,v,y',\vec{x}) & = & \int_0^\pi \frac{d \,k}{\pi/2}
\int_0^\pi \frac{d\,k'}{\pi/2} \int_\mathcal{B}
\frac{d\,\vec{q}}{(2 \pi)^{d-1}} \sin(k\,y) \sin(k' y')
e^{-\II \vec{x} \cdot \vec{q}} \hat{g}(t;u,k,v,k',\vec{q})
\EEA
Here, the sums are over all lattice sites, whereas $\mathcal{B} = [-\pi,\pi]^{d-1}$ 
is the first Brillouin zone and $d\, \vec{q} =
\prod_{i=1}^{d-1} d\,q_i$. It is straightforward to work out
equation (\ref{diffeqn_gf}) in Fourier space, which yields
\BEQ
\label{diffeqn_gf2}
\partial_t \hat{g}(t;u,k,v,k',\vec{q}) + \omega(k,k',\vec{q}) \;
\hat{g}(t;u,k,v,k',\vec{q};t) = \sin(u k) \sin(v k') \delta(t)
\EEQ
where the expression $\omega(k,k',\vec{q})$ is given by
\BEQ
\omega(k,k',\vec{q}) = \omega(k) + \omega(k') + \omega(\vec{q})
\EEQ
with
\BEQ
\omega(k) = \left(\frac{1}{d} - \gamma \cos(k)
\right), \quad \omega(k') =  \left(\frac{1}{d} - \gamma
\cos(k')\right), \quad \omega(\vec{q}) = 
\sum_{i=1}^{d-1}\left( \frac{2}{d} - 2 \gamma \cos(q_i) \right).
\EEQ
Equation (\ref{diffeqn_gf2}) is readily solved and yields
the result
\BEQ
\label{res_fouriers}
\hat{g}(t;u,k,v,k',\vec{r};t) = \Theta(t) \sin(u\,k)
\sin(v\,k') \exp\left(-\omega(k,k',\vec{q}) t\right)
\EEQ
where $\Theta(t)$ is the Heaviside step function.
This expression still has to be brought back to direct
space using (\ref{f_trafo2}). With the integral
$\int_{-\pi}^\pi d\,k \exp(\II r k + \cos(k) z) = 2
\pi I_r(z)$ this yields the result
\BEQ
\label{green_function}
g(t;u,y,v,y',\vec{x}) = \Theta(t) e^{-2 t} (I_{u-y}
(\gamma t) - I_{u + y}(\gamma t)) (I_{v-y'}
(\gamma t) - I_{v+y'}(\gamma t)) \prod_{i=1}^{d-1}
I_{x_i}(2 \gamma t).
\EEQ
With the help of this function, the inhomogeneous differential equation
(\ref{diffeq_onetimecorr}) is solved by
\BEQ
\label{res}
{\mathcal C}(t;y,y',\vec{x}) = {\mathcal C}_{h}(t;y,y',\vec{x}) + \sum_{u,v \geq 0}
\sum_{\vec{x}'} \int_0^\infty d\,\tau
g(t-\tau;u,y,v,y',\vec{x}-\vec{x}\,') b(\tau;u,v,\vec{x}')
\EEQ
where ${\mathcal C}_h(t;y,y',\vec{x})$ is an arbitrary solution of the
homogeneous equation $\square^{(\gamma)}_{t;y,y',\vec{x}}
\, {\mathcal C}_h(t;y,y',\vec{x}) = 0$ that satisfies the boundary
condition (\ref{boundary_cond}). 

In order to obtain this special solution
${\mathcal C}_h(t;y,y',\vec{x})$ we start from equation
(\ref{diffeqn_gf}) with a vanishing right hand side,
\BEQ
\square_{t;y,y',\vec{x}} \, {\mathcal C}_h(t;y,y',\vec{x}) = 0
\EEQ
for which we obtain the solution
\BEQ
\label{sp_sol1}
\hat{{\mathcal C}}_h(t;k,k',\vec{q}) = \exp\left(-\omega(k,k',\vec{q})
t \right) \hat{{\mathcal C}}_h(0;k,k',\vec{q})
\EEQ
in Fourier space
where the initial value $\hat{{\mathcal C}}_h(0;k,k',\vec{q})$ is given
by
\BEQ
\label{sp_sol2}
\hat{{\mathcal C}}_h(0;k,k',\vec{q}) = \sum_{u,v \geq 0}
\sum_{\vec{x}'(\vec{x})} \sin(u\,k) \sin(v\,k') e^{-\II
\vec{x}\;' \vec{q}}\, {\mathcal C}(0;u,v,\vec{x}\,').
\EEQ

The equation (\ref{diffeq_twotimecorr}) for the two-time correlator is also
solved in a similar way. We go to Fourier space and get
the solution
\BEQ
\hat{C}(t,s;k,k',\vec{q}) =
\exp\left(-\frac{\gamma}{2}\left(\omega(k)+\omega(\vec{q})\right)(t-s)\right)
\hat{{\mathcal C}}(s;k,k',\vec{q}).
\EEQ
Inserting the Fourier transform of the solution of equation
(\ref{res}), we obtain after transforming back to real space the final result
\BEA
 C(t,s;y,y',\vec{x}) &=& \sum_{u,v \geq 0}
\sum_{\vec{x}\;'(\vec{x})} {\mathcal C}(0;u,v,\vec{x}\;')\,
e^{-(t+s)}\prod_{i=1}^{d-1} I_{r_i-r'_i}(\gamma(t+s))
\nonumber \\ & \times &
\Big(I_{u-y}(\gamma t) - I_{u+y}(\gamma t)\Big)
\Big(I_{v-y'}(\gamma s) - I_{v+y'}(\gamma s) \Big)  \\
& + &\sum_{u,v \geq 0} \sum_{\vec{x}\;'(\vec{x})}
\int_0^\infty d\,\tau \,
b(\tau;u,v,\vec{x}') \, e^{-(t+s-2\tau)}
\prod_{i=1}^{d-1} I_{x_i-x'_i}(\gamma(t+s -2 \tau))
\nonumber \\ & &
\Big(I_{u-y}(\gamma(t-\tau)) - I_{u+y}(\gamma(t-\tau)) \Big)
\Big( I_{u-y}(\gamma(s-\tau)) - I_{u+y}(\gamma(s-\tau))\Big).
\nonumber
\EEA
For decorrelated initial conditions ${\mathcal C}(0;u,v,\vec{x}\;') =
\delta_{u,v}\delta_{\vec{x}\;',\vec{0}}$, one can rearrange
the Besselfunctions in the first sum using $\sum_{\nu =
-\infty}^\infty I_{\nu}(z_1) I_{\nu + k}(z_2) = I_k(z_1 +
z_2)$ which gives
\BEQ
\label{bessel_identity}
\sum_{u \geq 0} \Big(I_{u-y}(\gamma t) - I_{u+y}(\gamma t)\Big)
\Big(I_{u-y'}(\gamma s) - I_{u+y'}(\gamma s) \Big) =
I_{y-y'}(\gamma(t+s))-I_{y+y'}(\gamma(t+s)),
\EEQ
an equation needed for deriving the final result
(\ref{surf_corr_final}).

\end{document}